# Simplification Resilient LDPC-Coded Sparse-QIM Watermarking for 3D-Meshes

Bata Vasic*, Bane Vasic, *Fellow, IEEE*

*Abstract*— We propose a blind watermarking scheme for 3-D meshes that combines sparse quantization index modulation (QIM) with deletion correction codes. The QIM operates on the vertices in rough concave regions of the surface thus ensuring impeccability, while the deletion correction code recovers the data hidden in the vertices, which is removed by mesh optimization and/or simplification. The proposed scheme offers two orders of magnitude better performance in terms of recovered watermark bit error rate compared to the existing schemes of similar payloads and fidelity constraints.

## I. Introduction

DIGITAL watermarking, i.e. hiding a digital signal (watermark) into the original signal (cover signal or host signal) has been widely used for copy-protection of digital media. A good watermarking scheme must adequately balance the conflicting requirements of fidelity, and capacity, while ensuring robustness and security. Roughly speaking, fidelity referrers to the perceptual closeness of the watermarked cover signal to the original cover signal, and capacity is the amount of hidden data per bit of the host data. Security is a measure of inability of an adversary to alter or remove a watermark, and the robustness is the probability of detecting a watermark data under common operations on the watermark cover signal such as compression and affine transformations. Another common requirement is that watermarking is blind, i.e., the original cover signal is not required for watermark detection.

In the past decade, watermarking of audio, video and images has been widely studied and numerous techniques have been developed. On the contrary, blind watermarking of three-dimensional (3D) mesh objects is in its infancy, although the use of 3D models is growing rapidly. Watermark as copyright protection of these models is significant due to their mass use in virtual reality, modern computer games, and recently in digital cinematography as well as in computer-aided design (CAD), architectural and medical simulations. One of main difficulties in 3D watermarking is that even common signal processing operations for manipulation and editing 3D objects involve a number of complex geometric and topological operations very different from the transformations of signals in one dimension (1D) and two dimensions (2D). These transformations include mesh simplification and optimization, which are used to accelerate rendering. Specially, real-time 3D model rendering needs the significant geometry simplification, which is a serious threat to watermark data. While in essence analogous to compression of images using wavelets, mesh optimization and simplification involve much more complex "undersampling" resulting from removal of vertices and merging of polygonal faces in the original 3D mesh. The removal of a vertex from the watermarked cover 3D mesh may have a catastrophic effect to robustness: if the deleted vertex contains even a single bit of hidden data, not only is this bit undetectable, but its absence causes the loss of alignment between the original watermark and recovered watermark data, making the watermark extremely fragile if not unusable.

In this paper we propose a secure watermarking scheme with guaranteed robustness to mesh optimization and simplification based on error correction codes. We analyze a non-malicious transformation scenario, in which the only "irreparable" modification the mesh undergoes during the process of optimization is vertex deletion. It combines sparse quantized index modulation (QIM) for data hiding with run-length modulated low-density parity-check (LDPC) codes for recovering deleted watermark bits. The watermark recovery has two steps, namely detection of QIM bits [1] and error correction decoding. The redundancy (overhead) introduced in the coded watermark sequence is used in the decoding process to recover the deleted bits. Up to our best knowledge, this is the first error control coded scheme for watermarking of 3D meshes.

The rest of the paper is organized as follows. Section II describes the philosophy of our approach and its relation to prior work. Section III contains all necessary mathematical concepts and definitions, including QIM, LDPC codes and background on their iterative decoders. The proposed algorithm and detailed descriptions of its steps are given in Section IV. Numerical results of the selection of 3D mesh vertices designated for data hiding, as well as results of capacity estimation and error probability are given in Section V. Section VI gives discussion, conclusions and main directions of future research.

Manuscript received April 4, 2012, revised October 9, 2012. This work is funded in part by NSF, Grant CCF-0963726.

Bata Vasic is with the Department of Electronics, Faculty of Electronic Engineering Nis, University of Nis, 18000 Nis, Republic of Serbia (phone: +381-63-417-696; fax: +381-18-588-399; e-mail: bata.vasic@ elfak.ni.ac.rs).

Bane Vasic is with the Department of Electrical and Computer Engineering, University of Arizona, Tucson, AZ, 85721 USA (e-mail: vasic@ece.arizona.edu).

## II. Prior Work

Recently Coumou and Sharma [2] proposed a watermarking scheme for audio signal watermarking, which uses error correction coding. Their method is based on the insertion/deletion/substitution correction scheme of Davey and MacKay [3], which combines marker codes for providing synchronization and LDPC codes for error correction. To be effective in terms of correctable number of errors, the Davey and MacKay scheme must use long, non-binary LDPC codes, which make the decoding very complex. In addition, the algorithm involves using highly complex iterative maximum a posteriori (MAP) symbol detection operating on a huge graphical model of the insertion/deletion/substitution channel[1].

The philosophy of our approach is different. We employ strong signal processing during the watermark embedding to ensure proper bit synchronization prior to decoding, while the decoder is still binary and operates only on a code graph, not on channel trellis. The reasoning for such approach is the following. A mesh simplification process removes vertices, and aggressive watermarking, i.e. modifying a large number of vertices, results in the large number of deletion and substitution errors, thus requiring not only large data redundancy but also complex decoders. It is opposed to the state-of-the art methods which liberally allow occurrence of deletions, and then deal with them in a highly complex decoder.

Our content-aware embedding places the hidden data into the mesh vertices that stay invariant during the mesh simplification, thus reducing the probability of bit deletion. The heart of this method for selecting "stable" vertices is the *Ordered Statistics Vertex Extraction and Tracing Algorithm* (OSVETA) [4]. OSVETA is a sophisticated and powerful algorithm, which combines a number of mesh topology with human visual system (HVS) metrics to calculate vertex stabilities and trace vertices most susceptible for extraction during simplification. Such vertex preprocessing allows using low complexity decoding algorithms tailored to deletions, which is the dominant type of error during simplification.

This is achieved by another innovation proposed in this paper, which consists of protecting the watermarked bits by an LDPC code and modulating coded bits by a run-length code. As showed by Krishnan and Vasic [5] who use this idea in the context of data storage, the run-length code has a crucial role in transforming the notoriously difficult insertion/deletion/substitution channel which has infinite memory into a memoryless channel. Runlength coding introduces an inherent rate loss, but from the decoder side the transformed channel looks like the memoryless channel (such as *binary symmetric channel* (BSC)), and allows us to use the reach knowledge in LDPC codes to design powerful codes and low-complexity iterative decoders.

The LDPC codes and iterative decoders can be designed to guarantee upper bounds of watermark decoding probability or the *frame error rate* (FER) for a *code rate*. Code rate is the ratio between the uncoded and coded watermark lengths, and is upper-bounded by the Shannon capacity of the channel which is known for memoryless channels. This is in contrast with design of codes for channels with infinite memory for which theoretical bounds for achievable code rates are unknown, which practically means that a watermarking system designer does not know a priory how to choose a code suitable to the channel.

The data embedding component of the method employed in this paper is based on the QIM [6]. In the QIM each bit of the watermark sequence is encoded by the choice of one of two uniform quantizers that are applied to the cover signal. The watermarking robustness and fidelity is controlled by the step size of the underlying dithered uniform quantizer. A larger step leads to more robust detection, but reduces the cover signal fidelity. We use the sparse variant of the QIM proposed by Chen and Wornell [1], also known as *spread-transform dither modulation*, which was shown to be provably secure [6].

Although the definitions of QIM and *sparse-QIM* (s-QIM) do not put any restrictions on the type of a cover signal, their application to 3D-meshes requires addressing the nontrivial issues connected to robustness to common transformations. Recently Darazi, *et al.* [7] adapted the sparse QIM (s-QIM) for 3D-mesh watermarking. To ensure robustness to affine transformation, quantization is performed in the spherical coordinate system and the distance from the center of gravity (mass) is the only coordinate that is quantized. To ensure adequate fidelity as well as robustness to mesh simplification, watermark data is not hidden in all vertices but only those vertices in the areas with high curvature and roughness because such vertices are not removed by the simplification algorithms. Improving watermark robustness by exploiting the masking effect of surface roughness on watermark visibility is a feature implemented in two popular algorithms by Benedens [8] and by Cayre and Macq [9]. See also recent work by Kim *et al.* [10] for an advanced version of these algorithms and an overview of literature in this area.

In this paper we consider the scenario in which the distortion of the signal occurs only due to common, non-malicious mesh simplifications. As we see, in this case, the data hiding channel introduces deletions only.

## III. Preliminaries

In this section, we introduce the notation and the concepts used through the paper. We start with a brief discussion of the embedding and detection in QIM and sparse QIM. Then we introduce quantization of polar coordinates, and concept of Hausdorf distance, roughness, and curvature. We also introduce basic terminology of LDPC codes and iterative decoding as well as the runlength coding.

### A. Quantized Index Modulation

Let $\mathbf{u} \in \{0,1\}^n$ and $\mathbf{x} \in \mathbf{R}^n$ be the watermark sequence and the cover sequence, respectively. The *embedder* combines the $n$-dimensional vectors $\mathbf{u}$ and $\mathbf{x}$ and produces the watermarked sequence $\mathbf{y} \in \mathbf{R}^n$. The difference $\mathbf{w}=\mathbf{y}\text{-}\mathbf{x}$ is referred to as the *watermarking displacement* signal. The embedder must keep the distortion $d(\mathbf{x},\mathbf{y})$ within a prescribed limit, i.e., $d(\mathbf{x},\mathbf{y}) \leq nD$, where $D$ is the maximum allowed distortion per dimension for every $\mathbf{x}$ and $\mathbf{u}$. The distortion is typically defined as the simple Euclidian distance or the Hausdorff distance which is more suitable to the HSV. The Hausdorff distance is defined as

$$d_{Hausdorff}(\mathbf{x},\mathbf{y}) = \max\left\{\sup_{x\in\mathbf{x}}\inf_{y\in\mathbf{y}} d(x,y), \sup_{y\in\mathbf{y}}\inf_{x\in\mathbf{x}} d(x,y)\right\}. \quad (1)$$

---

[1] The watermark recovery also includes decryption, but the cryptography aspect of watermarking is beyond scope of this paper. In the rest of the paper it is assumed that watermark data are encrypted by an outer encryption system.

The QIM operates on independently on the elements $u$ and $x$ of the vectors $\mathbf{u}$ and $\mathbf{x}$. To embed the bit $u\in\{0,1\}$, the QIM requires two uniform quantizers $\mathbf{Q}_0$ and $\mathbf{Q}_1$ defined as the mappings

$$\mathbf{Q}_u(x) = \Delta\left[\frac{1}{\Delta}\left(x - (-1)^u \frac{\Delta}{4}\right)\right] + (-1)^u \frac{\Delta}{4}, \quad (2)$$

where, with a slight abuse of notation, [ ] denotes the rounding operation, i.e. for a real $x$, $[x]$ is the integer closest to $x$. Thus, the quantization level of the "nominal" quantizer $\Delta [x/\Delta]$ is moved up or down by $\Delta/4$ depending on the value of $u$. Equivalently, the watermark bit $u$ dithers the input $x$ by the amount $\pm\Delta/4$. The watermark bit $u$ determines the selection of a quantizer, so that $y = \mathbf{Q}_u(x)$.

In the simplest case of the AWGN attack, the received signal is $\mathbf{w}=\mathbf{y}+\mathbf{e}$ where $\mathbf{e}$ is the noise vector introduced by the channel (a malicious user). The recovered bit is calculated as

$$\hat{u} = \arg\min_{u\in\{0,1\}} \|w - \mathbf{Q}_u(w)\|_2 \quad (3)$$

where $\|\ \|_2$ denotes the Euclidean or $l_2$ norm. If the channel introduces deletions, then the received sequence has the form

$$\left(y_1, y_2, ..., y_{i_1-1}, y_{i_1+1}, ..., y_{i_2-1}, y_{i_2+1}, ..., y_{i_l-1}, y_{i_l+1}, ..., y_{n-l}\right) \quad (4)$$

where $i_1, i_2, ..., i_l$ are the positions of $l$ deletions. We will also use the symbol $\sqcup i$ to denote that $i$-th bit is deleted. The minimum error produced by QIM is $\Delta/2$. Assuming uniform distribution of the quantization errors over the interval $[-\Delta/2, \Delta/2]$, the mean square error distortion is $\Delta^2/12$.

### B. Sparse QIM

The QIM is independent of the 3D-object content, and since a small error in the Euclidean sense may be perceived by the HVS as a large distortion, the general version of the QIM may lead to serious degradation of the quality of the watermarked 3D signal. To avoid this problem, the vector $\mathbf{x}$ is chosen so that the application of QIM does not change visual quality.

Sparse QIM, spreads out the watermark bit over $L$ elements of cover signal $\mathbf{x}$. The cover sequence $\mathbf{x}_L$ of length is projected to a $L$-dimensional vector $\mathbf{p}$ of the unit norm, and the norm of the corresponding projection is quantized. The resulting $L$-dimensional watermarked vector $\mathbf{y}_L$ can be written as

$$\mathbf{y}_L = \mathbf{x}_L + \left(\mathbf{Q}_u(\mathbf{x}_L^T\mathbf{p}) - \mathbf{x}_L^T\mathbf{p}\right)\mathbf{p} \quad (5)$$

The detector projects the received watermarked cover vector $\mathbf{r}_L$ to $\mathbf{p}$ and recovers the embedded bit as

$$\hat{u} = \arg\min_{u\in\{0,1\}} \|\mathbf{r}_L^T\mathbf{p} - \mathbf{Q}_u(\mathbf{r}_L^T\mathbf{p})\|_2. \quad (6)$$

Generally, the robustness of the watermark increases with $L$. The choice of the masking vector $\mathbf{p}$ is explained in subsequent sections discussing curvature and roughness.

### C. QIM in Spherical Coordinates to Increase Robustness to Affine Transforms

Up to this point, we have treated the cover signal QIM in a rather general way. The cover signal was a vector of certain length, and the QIM operated on components of this vector. In 3D watermarking however, the cover signal consists of mesh vertices in the three-dimensional space, and HSV imposes restrictions on the QIM embedding design. In this subsection we introduce the coordinate system changes to make the watermarked object robust to common affine transformations.

Following [11], to ensure invariance to translation and rotation, the coordinate system is changed by translating the coordinate origin to the mass center and by aligning the principal component vector with the $z$ axis. The principal component axis is the eigenvector that corresponds to the largest eigenvalue of the vertex coordinate covariance matrix.

To achieve robustness to scaling, the Cartesian coordinates of the point are then converted to spherical coordinates. The original variant of the QIM operates only on the radial distance from the center of mass. The above conditions are satisfied in practice, and the watermarked object is therefore invariant to affine transformations.

### D. Low Density Parity Check Codes

Here we provide some definitions related to LDPC codes. For more information the reader is referred to [12]. Let $\mathcal{C}$ denote an $(n, k)$ LDPC code over the binary field GF(2). $\mathcal{C}$ is defined by the null space of $H$, an m × n parity-check matrix of $\mathcal{C}$. $H$ is the bi-adjacency matrix of $G$, a Tanner graph representation of $\mathcal{C}$. $G$ is a bipartite graph with two sets of nodes: $n$ variable (bit) nodes $V = \{1, 2,...,n\}$ and $m$ check nodes $\mathcal{C} = \{1, 2; ...;m\}$. A vector $\mathbf{x} = (x_1, x_2,...,x_n)$ is a codeword if and only if $\mathbf{x}H^T = 0$, where $H^T$ is the transpose of $H$. The support of $\mathbf{x}$, denoted as supp($\mathbf{x}$), is defined as the set of all variable nodes (bits) $v\in V$ such that $x_v \neq 0$. A $d_v$-left-regular LDPC code has a Tanner graph $G$ in which all variable nodes have degree $d_v$. Similarly, a $d_c$-right-regular LDPC code has a Tanner graph G in which all check nodes have degree $d_c$. A $(d_v; d_c)$-regular LDPC code is $d_v$-left-regular and $d_c$-right-regular. Such a code has rate R≥1-$d_v/d_c$ [13]. The degree of a variable node (check node, resp.) is also referred to as the left degree (right degree, resp.) or the column weight (row weight, resp.). The length of the shortest cycle in the Tanner graph $G$ is called the girth $g$ of $G$.

### E. The Sum-product Algorithm)

For any codeword $x = (x_v)_{1\leq v\leq n}$ in a linear block code given by the parity check matrix $H$, the following set of equations is satisfied

$$\sum_v h_{c,v} x_v = 0 \quad (7)$$

for $1\leq c\leq m$. The above equations are called parity check equations. Iterative decoding can be visualized as message passing on a Tanner graph [14]. There are two types of vertices in the graph: check vertices (check nodes) indexed by $c$ and variable vertices (bit nodes) indexed by $v$. An edge connecting vertices $c$ and $v$ exists if $h_{c,v}=1$, i.e. if variable $v$ participates in the parity check equation $c$.

The sum-product algorithm maybe described as follows. First, a priori information of the bit at position $j$, $\mu_j^{(0)}$, is taken as the log-likelihood ratio of the $i$-th bit, $\log(P(x_i = 0 | r_i) / P(x_i = 1 | r_i))$, where $P(x_i = 0 | r_i)$ is the a posteriori probability of the bit $x_i$ being zero given the received symbol $r_i$. $P(x_i = 1 | r_i)$ is defined analogously. The messages passed from variable node $j$ to check node $c$ in the bipartite graph, $\lambda_{j,c}^{(0)}$, are initialized to $\mu_j^{(0)}$. In $i$-th iteration we update the messages to be passed from check node $c$ to bit node $j$, $\Lambda_{c,j}^{(i)}$, as

$$\Lambda_{c,j}^{(i)} = -2 \tanh^{-1}\left( \prod_{w \neq j} \tanh\left( \lambda_{w,c}^{(i-1)} / 2 \right) \right) \quad (8)$$

and messages to be passed from bit node $j$ to check node $c$, $\lambda_{j,c}^{(i)}$, according to:

$$\lambda_{j,c}^{(i)} = \mu_j^{(0)} + \sum_{d \neq c} \Lambda_{d,j}^{(i)}. \quad (9)$$

The last step in iteration $i$ is to compute updated log-likelihood ratios $\mu_j^{(i)}$ according to:

$$\mu_j^{(i)} = \mu_j^{(0)} + \sum_c \Lambda_{c,j}^{(i)}. \quad (10)$$

For each bit $j$ the estimation is made according to

$$\hat{x}_j = \begin{cases} 1, & \text{if } \mu_j^{(i)} < 0 \\ 0, & \text{otherwise} \end{cases}. \quad (11)$$

The procedure halts when a valid codeword is generated or a maximum number of iteration has been reached.

## IV. THE PROPOSED ALGORITHM

The block diagram of the proposed watermarking system is shown in Fig. 1. In this section we present the method for robust blind watermarking based on LDPC codes. The subsections explain the essential blocks of the system.

One of the difficulties in addressing the problem of code construction is the fact that channels with synchronization errors have infinite memory, i.e., a synchronization error affects all subsequent symbols. In this paper, we use a coding scheme which enables to transform certain channels with synchronization errors into memoryless channels. In the next section we explain this encoding, called *runlength coding*.

### A. Runlength coding

In order to transform the channel with infinite memory into a memoryless channel, the bits (symbols) at the input of the channel are represented by runs of bits. The runlenghts representing zeros and ones are selected according to channel synchronization statistics. To explain the main idea, let us consider a case in which binary zeros are represented by runs of length *two*, and binary ones with runs of length *three*.

As an illustration, consider the sequence of information bits $b = (0, 1, 1, 0, 1, 1)$. The encoded sequence $c$ is initialized to an empty string. The encoding proceeds as follows.

The first bit $b_1$ is 0. To encode $b_1$, a run consisting of *two* bits, namely $c_1$ through $c_2$, is added to $c$. This can be done by setting $c_1 = c_2 = 1 \mod 2 = 1$ ($c = (11)$). Next, since $b_2$ is 1, we construct a run consisting of three bits, namely, $c_3$ through $c_5$, is added to $c$. This can be done by setting $c_4 = c_5 = c_6 = 2 \mod 2 = 0$ ($c = [11\,000]$). Proceeding thus, we find that $b$ is encoded as encoded bits $c = (11\,000\,111\,00\,111\,000)$.

In other words, input symbols with odd indices are encoded in runs of $k$ 1's, with $k = 2$ for symbol 0 and $k = 3$ for symbol 1. Symbols with even indices are similarly encoded by runs of 0's.

As shown in Fig. 1., the runlength sequence is embedded into selected vertices of a 3D object by using sparse-QIM.

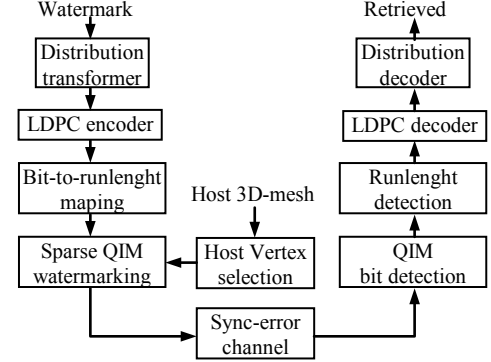

Fig. 1. A block diagram of the proposed LDPC-coded watermarking system

However, prior to conversion to runs and transmission through the channel, the watermark binary sequence is first processed in the distribution transformer to optimize probability of zeros and ones, and encoded by a binary LDPC code.

### B. Distribution Transformer

The reasons for optimizing probability of symbols comes from the fact that symbols have different lengths. The Shannon capacity of memoryless channels is obtained by maximizing the mutual information, $I(\mathbf{X}; \mathbf{Y})$, between the input alphabet $\mathbf{X}$ and the output alphabet $\mathbf{Y}$, over all input distributions of $\mathbf{X}$.

The Shannon capacity of a discrete memoryless channels is obtained by maximizing the mutual information, $I(\mathbf{p})$, over all input probability distribution vectors $\mathbf{p} = (p(x))_{x \in X}$. Formally, Shannon capacity of a channel $C$ is defined as $C = \max_{\mathbf{p}} I(\mathbf{p})$ where

$$I(\mathbf{p}) = \sum_{x \in X} \sum_{y \in Y} p(x) p(y | x) \log \frac{p(x) p(y | x)}{p(x) \sum_{x \in X} p(x) P(y | x)} \quad (12)$$

and $P = [p(y | x)]_{x \in X, y \in Y}$ is the channel transition probability matrix.

This definition assumes that the costs of transmission of different symbols are equal. However, this is not true for the channel where symbol lengths are not equal. Let $c(x)$ denote the length of symbol $x$. More generally, $c(x)$ can be viewed as the const of transmission of the symbol $x$. Let define $\mathbf{c}$, the transmission cost vector, $\mathbf{c} = (c(x))_{x \in X}$. In such channels, we use the notion of *unit-cost capacity* [14].

$$C_{unit} = \max_{\mathbf{p}} \frac{I(\mathbf{p})}{\mathbf{c} \mathbf{p}^T} \quad (13)$$

In our example the cost of transmitting the information-bit 0 is 2, since 2 channel bits are used for transmission. Similarly, the cost of transmitting the information-bit 1 is 3.

It can be seen that in our transmission model, the unit-cost capacity translates to capacity per channel-use for the transmission scheme under consideration. The capacity is a function of the synchronization error probability as well as the size of the input alphabet **X**. By changing the alphabet-size at the input, a spectrum of transmission schemes can be obtained. In general, for a channel $C$, with information alphabet **X**, we can calculate an associated unit-cost capacity $C(C; |\mathbf{X}|)$, and determine optimal capacity-achieving input probabilities. The capacity achieving distribution $p(\mathbf{x})$ can be obtained by Jimbo-Kunisawa algorithm [15]. The distribution transformer is readily achieved by employing arithmetic decoder, which is widely used in source coding applications [16]. The data is then encoded by using an error-correcting code. Subsequently, the information symbols are encoded in runs of channel bits as described previously. At the output, all the operations are performed in reverse in order to obtain an estimate of the input data.

*C. LDPC Coding*

In this section, we give a description of structured LDPC codes used in this paper. The parity-check matrices of these codes are arrays of permutation matrices obtained from Latin squares. As we showed in [17], this class of codes has excellent performance in variety of channels.

A permutation matrix is a square binary matrix that has exactly one entry 1 in each row and each column and 0's elsewhere. Our codes make use of permutation matrices that have disjoint support. These sets of permutation matrices can be obtained conveniently from Latin squares. A Latin square of size $q$ (or order $q$) is a $q \times q$ array in which each cell contains a single symbol from a $q$-set $S$, such that each symbol occurs exactly once in each row and exactly once in each column. A Latin square of size $q$ is equivalent to the Cayley table (or multiplication table) of a quasi-group $\mathcal{Q}$ on $q$ elements (see [17] for details). For mathematical convenience, we use elements of $\mathcal{Q}$ to index the rows and columns of Latin squares and permutation matrices.

Let $\mathcal{L} = \left[l_{i,j}\right]_{i,j \in \mathcal{Q}}$ denote a Latin square defined on the Cayley table of a quasi-group $(\mathcal{Q}, \otimes)$ of order $q$. We define $f$, an injective map from $\mathcal{Q}$ to Mat($q, q$, GF(2)), where Mat($q, q$, GF(2)) is the set of matrices of size $q \times q$ over GF(2), as follows:

$$f: \mathcal{Q} \to \text{Mat}(q, q, \text{GF}), \quad \alpha \to f(\alpha) = \left[m_{i,j}\right] \quad (14)$$

such that

$$m_{i,j} = \begin{cases} 1 & \text{if } l_{i,j} = 0 \\ 0 & \text{if } l_{i,j} \neq 0 \end{cases}. \quad (15)$$

According to this definition, a permutation matrix corresponding to the element $\alpha \in \mathcal{Q}$ is obtained by replacing the entries of $\mathcal{L}$ which are equal to $\alpha$ by 1 and all other entries of $\mathcal{L}$ by 0. It follows from the above definition that the images of elements of $\mathcal{Q}$ under $f$ give a set of $q$ permutation matrices that do not have 1's in common positions. This definition naturally associates a permutation matrix to an element $\alpha \in \mathcal{Q}$ and simplifies the derivation of parity-check matrices that satisfy the constraint of absence of four cycles in the corresponding Tanner graph.

Let $W$ be a matrix $\mu \times \eta$ over a quasi-group $\mathcal{Q}$, i.e.,

$$W = \begin{bmatrix} w_{1,1} & w_{1,2} & \cdots & w_{1,\eta} \\ w_{2,1} & w_{2,2} & \cdots & w_{2,\eta} \\ \vdots & \vdots & \ddots & \vdots \\ w_{\mu,1} & w_{\mu,2} & \cdots & w_{\mu,\eta} \end{bmatrix}. \quad (16)$$

With some abuse of notation, let $\mathcal{H} = f(W) = \left[f(w_{i,j})\right]$ be an array of permutation matrices obtained by replacing elements of $W$ with their images under $f$, i.e.,

$$\mathcal{H} = \begin{bmatrix} f(w_{1,1}) & f(w_{1,2}) & \cdots & f(w_{1,\eta}) \\ f(w_{2,1}) & f(w_{2,2}) & \cdots & f(w_{2,\eta}) \\ \vdots & \vdots & \ddots & \vdots \\ f(w_{\mu,1}) & f(w_{\mu,2}) & \cdots & f(w_{\mu,\eta}) \end{bmatrix}. \quad (17)$$

Then $\mathcal{H}$ is a binary matrix of size $\mu q \times \eta q$. The null space of $\mathcal{H}$ gives an LDPC code $\mathcal{C}$ of lenght $\eta q$. The column weight and row weight of $\mathcal{C}$ are $d_v = \mu$ and $d_c = \eta$, respectively.

*D. Dealing with Roughness and Curvature in QIM*

The vertex *importance* is proportional to the curvature of its surrounding surface. The removal of these *important* vertices destroys information about the shape, so simplification processes avoid those vertices. Nevertheless, for large threshold values of simplification some of these vertices are deleted. These are the vertices in the areas of lower curvature. Vertices with the highest curvature remain stabile even under the most aggressive decimation. For watermark embedding we need *stabile* vertices of a given 3D mesh. From the data hiding point of view, a vertex is called stabile when its coordinates remain unchanged under the process of simplification. The idea of OSVETA is to select vertices with the highest curvature, and use certain topological and HVS metrics to create the vertex stability ordered statistics. For more details of OSVETA the reader is referred to [4].

The input of OSVETA is $M(V,F)$, a mesh of a given 3D surface. The essence of OSVETA represent three steps: defining assessment criteria and their ranking, accurate curvature evaluation with its characteristic features computation, and tracing importance of extracted vertices in relation to mesh topology. In the first step the algorithm extracts the matrix of topological error vertices [4] and the matrix of boundary vertices. Using the values and signs of Gaussian $\kappa_G(\mathbf{v}_i)$ and mean curvature $\kappa_H(\mathbf{v}_i)$ at vertex $\mathbf{v}_i$, algorithm extracts *risky* vertices. Then, using other features, the algorithm calculates vertex stabilities. The result of OSVETA are two vectors: **s**, the vector of vertex stabilities arranged in a decreasing order, and **i**, the vector of corresponding vertex indices. The mesh vertices that are ordered with respect of decreasing stability form the vector $V_o$.

$V_o=\mathbf{v_i}$, and the length-$L$ vector **p** for watermark embedding are obtained by taking the first $L$ elements from **i**.

The OSVETA is practically implemented so that a user selects the desired watermark length, and based on the object topology, the vertex selection criteria are determined. Then vertices are selected or rejected with respect to the given curvature criteria. The criteria may be tightened to extract only the vertices from rough or concave regions of surface. This significantly increases the success of data hiding, and reduces a perceptual visibility of watermark. Since the number of stable vertices depends on mesh topology, there is an optimal watermark length for each object. However, determining such optimum is nontrivial and is beyond scope of this paper.

*E. Synchronization Error Channel*

As mentioned in the introductory remarks, in this section, the runlength coding naturally leads to a scheme in which conventional error-correcting codes may be used to compensate for insertion/deletion errors. In this section, based on our work in [5], we give a brief description of the system model synchronization-error channel we consider, and explain how information may be reliably transmitted through this channel.

Consider a channel with binary input and output alphabets. The most general formalism that describes synchronization and bit-flip error is a finite-state machine [3]. At time $t_i$ a bit may be inserted with a probability $p_i$, deleted with a probability $p_d$, or successfully transmitted with a probability $1-p_i-p_d$. We refer to insertions and deletions as to *synchronization errors*. The possibility of a change in the value of the transmitted bit can also be encompassed by this model.

In the context of watermarking, the channel introduces only deletions ($p_i = 0$). Furthermore we assume that no two consecutive vertices are deleted in the process of simplification. This assumption is justified by the fact that the OSVETA selects stable vertices, and two consecutive vertices are deleted extremely rarely. In other words, the space between two consecutive deleted vertices is sufficiently large, i.e., there are no *bursts* of synchronization errors. Separation of synchronization errors within a codeword may be also achieved by interleaving, i.e., by proper indexing of vertices. Consequently, we consider a variant of the channel described above in which the number of consecutive synchronization errors is restricted and operates on individual runs of binary sequences. Here, we define the term *run* in a binary sequences as an occurrence of $k$ consecutive, identical symbols.

We now define a channel with the parameter $s_d$, which is the maximum number of consecutive deletions. We denote this channel $C$ as ($p_d$, $s_d$). It is a special case of the channel we introduced in [5]. The probability of $j$ consecutive deletions is $(p_d)^j$ for $j = 1,...,s_d$. In each run of zeros or ones, only one of the $s_d + 1$ error events occurs, namely, (i) error-free transmission, or (ii) deletion of $s_d$ bits, $j = 1,...,s_d$. We do not consider errors in bit-values introduced by channel since we consider only common transformations. In general, these parameters may be chosen to match the behavior of the object in which we hide the watermark.

To further illustrate the behavior of the channel, consider as example a channel with parameters ($p_d$, 2). Let $b_l$ denote to the $l^{th}$ bit with value $b \in \{0, 1\}$. Let $\sqcup_l$ represent a deleted bit at position $l$ (note that the receiver cannot identify the position or value of a deleted bit. These indices are for ease of illustration only). Suppose that the sequence ($0_1$, $0_2$, $0_3$, $0_4$, $0_5$) is transmitted through this channel. Then, the sequence ($0_1$, $\sqcup_2$, $0_3$, $0_4$, $0_5$) is an example of a valid output since there are no more than two consecutive deletions. However, ($0_1$, $\sqcup_{2i}$, $\sqcup_3$, $\sqcup_4$, $0_5$), is not a valid output since there are more than 2 consecutive deletions.

Consider the channel $C$ with parameters ($p_d$, $s$), having input **x**. As we have shown in [5] if all the runs in **x** have lengths greater than $s$, then the number of runs in any output **y** produced by $C$ is equal to the number of runs in **x**. Summarizing, when transmitted through $C$, each run of $s_t$ bits, $s_t > s$, results in a run of $j$ bits, $j = s_t -s,..., s_t$. This observation leads naturally to an encoding scheme where symbols are encoded in runs of bits. To explain this, consider a channel with $s_d = 1$. At the receiver, the length of each run is counted in order to determine the output symbol. We note that at the receiver, the lengths of runs may change due to deletion. Nevertheless, a correct choice of runlengths (more precisely, by choosing runlengths greater than $s$) assigned to information symbols can lead to the $i^{th}$ information symbol corresponding exactly to the $i^t$ run of channel bits.

As we noted in [5], an important consequence of such an encoding scheme is that transmission through the synchronization-error channel can now be represented as that of transmission through a memoryless channel. That is, any synchronization-error manifests as an error in the corresponding output symbol, and does not affect other symbols. Thus, classical error-correcting codes can be used to compensate for such errors.

The encoding scheme described above results in a discrete memoryless channel as illustrated in Fig. 2.(a). The input symbols $l_0$ and $l_0+1$ denote the runlengths corresponding to the input bits 0 and 1. The output of the channel correspond to the all the possible values of runlengths at the receiver. It is easy to see that this methodology is not restricted to encoding of binary-valued sequences, and can be easily extended to larger alphabets. Fig. 2.(b) shows an example of transmission of information with alphabet sized four. In Fig. 2.(b) an input sequence is grouped into pairs of symbols (00, 01, 10 and 11), and each pair is represented by a unique runlength (00 by runlength 2, 01 by a runlengt 3 etc.). For example, an input sequence 0101001011 is parsed as 01, 01, 00, 10, 11 and then runlength encoded to obtain the sequence 000,111,00,1111,00000.

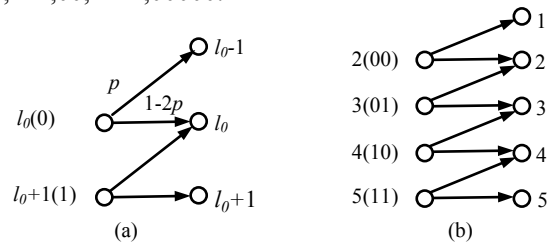

Fig. 2. An illustration of (a) discrete memoryless channel (b) example of transmission of information with alphabet sized four

## V. NUMERICAL RESULTS

### A. OSVETA Performance

For computing vertex stability in relation with optimization process we used 4 mesh models: (A) Christ the Redeemer [18], (B) Myron of Eleutherae [18], (C) Naissa by Bata [19], and (D) Venus de Milo [18]. For brevity we refer to these objects as to A, B, C, and D. These four models differ in total number of vertices, (i.e. faces) but also in their geometric structure. More precisely, they differ with respect to the percentage of curved and flat areas. Myron of Eleutherae and Naissa by Bata are complex mesh models; both containing closed elements as subobjects. OSVETA algorithm works equally well with both homogeneous and complex meshes, as well as with open meshes with a boundary.

For comparison and evaluation we used 'Optimize' modifier from 3D Studio Max 2012 application [20]. We have performed 7 different levels of optimization with the following face thresholds (FT) for the objects A, B, C and D:

$FT_A=(0,2,4,6,8,10,13,25)$, $FT_B=(0,2,4,6,8,10,13,27)$,
$FT_C=(0,2,4,6,8,10,13,26)$, $FT_D=(0,2,4,6,8,10,13,24)$.

The optimization with the maximum FT value completely destroys the geometric structure of the mesh, leaving only 5-10% of the total number of vertices. Higher maximal values are used for meshes with the larger total number of faces. Determination of face threshold limits preserving the usability of a 3D mesh is illustrated in Fig. 3. on the example of the Venus de Milo mesh.

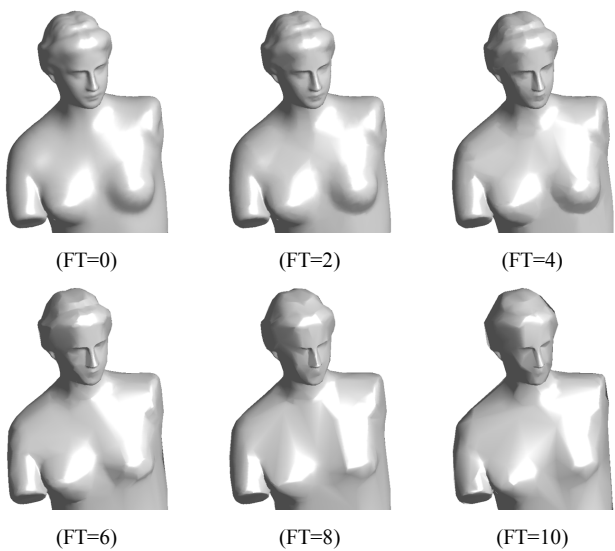

Fig. 3. Perceptual degradation of the Venus de Milo 3D mesh as a function of the face threshold level. The subfigures correspond to the threshold levels 0,2,4,6,8,10, respectively.

As can be seen from TABLE I, the optimization face threshold higher then 8, leads to significant perceptual and geometric degradation of mesh. Thus we therefore set the OSVETA criteria [21] for assessing vertices without considering FT optimization values over 10.

Experimental tests of stability 1000 vertices, selected by OSVETA algorithm compared to the randomly allocated group of 1000 vertices showed the superiority of our approach compared to random selection of vertices. The results are summarized in TABLE I.

TABLE I
THE NUMBER OF VERTICES DELETED BY SIMPLIFICATION, USING RANDOM AND OSVETA SELECTION ALGORITHMS.

| 3D Model | FT=0 | FT=2 | FT=4 | FT=6 | FT=8 | FT=10 |
|---|---|---|---|---|---|---|
| A#vertices | 27802 | 16627 | 9789 | 5991 | 3743 | 2432 |
| Random | 0 | 502 | 718 | 838 | 907 | 948 |
| OSVETA | 0 | 8 | 51 | 170 | 333 | 531 |
| B#vertices | 100681 | 49931 | 32353 | 21498 | 15078 | 11620 |
| Random | 0 | 523 | 686 | 794 | 848 | 877 |
| OSVETA | 0 | 6 | 8 | 19 | 51 | 94 |
| C#vertices | 33465 | 25334 | 17472 | 12146 | 8588 | 5870 |
| Random | 0 | 276 | 495 | 654 | 760 | 845 |
| OSVETA | 0 | 3 | 21 | 44 | 77 | 156 |
| D#vertices | 17350 | 12209 | 6953 | 3926 | 2315 | 1448 |
| Random | 0 | 332 | 622 | 781 | 872 | 920 |
| OSVETA | 0 | 1 | 30 | 147 | 332 | 522 |

The first row in each sub-table gives the total number of vertices remaining after simplification with a given FT. The second and third row gives the number of removed vertices out of 1000 vertices selected randomly and selected by the OSVETA. The nonzero numbers in the FT=0 column correspond to the total number of vertices in the original object.

The probability of vertex detection as a function of FT is shown in the Fig. 4. It is obtained as a ratio of the number of deleted edges and the total number of edges after simplification.

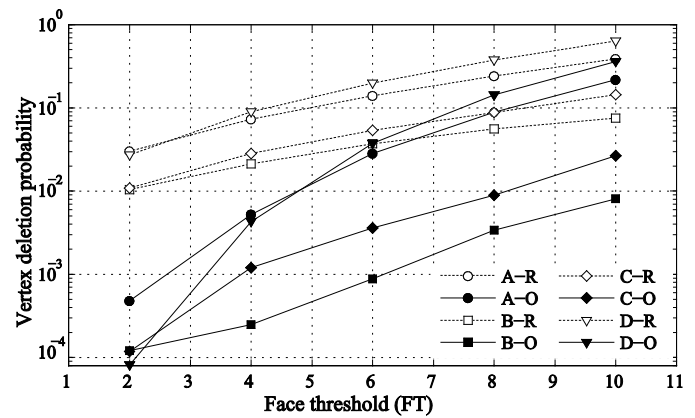

Fig. 4. Probability of vertex deletion $p_d$ as a function of the face threshold (FT) for the random (R) selection of vertices (white markers), and vertices selected by OSVETA (O - black markers) for the four objects: (A) Christ the Redeemer, (B) Myron of Eleutherae, (C) Naissa by Bata, and (D) Venus de Milo

It clearly shows that OSVETA provides two orders of magnitude better probably of deletion compared to random selection of vertices. For example, simplification of Myron of Eleutherae (mesh B) with FT=2, FT=4 and FT=6 removes 523, 686 and 794 out of 1000 vertices selected randomly, which renders the random selection useless, since no error correction system can handle such high number of deletions. On the contrary, the number of vertices removed from the set selected by the OSVETA is only 6, 8 and 19, which is quite readily tractable by the proposed codes. Note that 19 deleted vertices is an insignificant number in relation of 79183 total deleted vertices for FT = 6. The important fact is that the OSVETA provides an extremely low probability of deleting two consecutive selected vertices. Actually, in 1000 selected vertices of all meshes and also all of optimization levels optimization we did not find any deleted pair of consecutive vertices. This justifying to the ($p$,1) deletion channel assumption.

### B. Capacity Estimates

In order to determine bounds on the achievable rate of error correcting codes that may be used, we need to calculate the capacity of these channels. In this subsection, we investigate the capacity limits of channels described above. For a channel $C$, with information alphabet $\mathbf{X}$, we can calculate an associated unit-cost capacity $C(C; |\mathbf{X}|)$. These capacities constitute lower bounds on the capacity of $C$.

As an illustration, we give results for the channel: ($p$, 1) where Fig. 5. shows the capacity estimates of this channels for various values of $p$ and alphabet size $|\mathbf{X}|$ obtained by the Jimbo-Kunisawa algorithm [15]. The region of low channel deletion probability is of practical interest as in this case the watermark can be efficiently protected by a code.

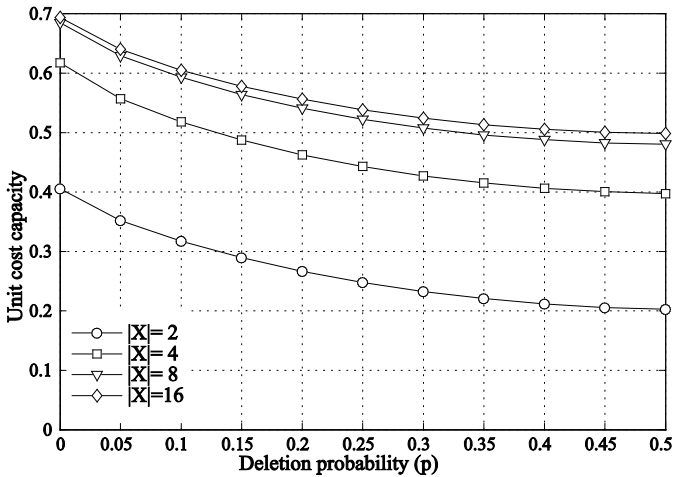

Fig. 5. Capacity estimates for the channel ($p$, 1) The alphabet sizes are in the set {2, 4, 8, 16}. For alphabet size |X|, the costs of transmission of a bit is in the set {2,…|X| + 1}. The capacity increases with the size of the alphabet.

As the symbol costs (runlengths) are not uniform, the maximizing input distribution dictates that symbols with low transmission-cost have a higher probability than those with high cost.

### C. Probability of Error Performance

In order to demonstrate the feasibility of implementation of our encoding methodology, we conducted experiments to simulate transmission of coded information through the channel ($p$, 1). For our simulations, we chose synchronization-error probability ($p$) ranging from 0.01 to 0.05. Two codes were chosen as candidates for simulation. The codes were constructed by methods described in [16], and provide guarantees on error-correction capability under iterative error-correction decoding for BSC. The code parameters are given in TABLE II. $R_{eff}$ denotes the effective rate, i.e. the number of information bits transmitted per channel use. For simplicity, in the simulations the equiprobable inputs are used. Since the average cost of a bit transmission is 2.5 (symbols of duration 2 and 3 are used half of the time in average), there runlength coding introduces a rate loss of 2/5. The effective rate is thus $R_{eff}=(2/5)R$ where $R$ is the rate of the LDPC code.

TABLE II
THE LDPC CODE PARAMETERS.

| | | | | | |
|---|---|---|---|---|---|
| Code 1 | $n$=2212 | $k$=1899 | $\eta$=361 | $R$=0.86 | $R_{eff}$=0.34 |
| Code 2 | $n$=848 | $k$=661 | $\eta$=186 | $R$=0.78 | $R_{eff}$=0.32 |

The code length is denoted by $n$, $k$ is the number of message bits in the codeword, $\eta$ is size of a block in the parity check matrix, $R$ is the code rate, and $R_{eff}$ is the effective runglengh coding rate.

Figure Fig. 6. shows the bit-error rates (BER) and frame-error (FER) rates with respect to the vertex deletion probability parameter $p$. Sum-product algorithm was used for decoding the codes. As can be seen, even with short- and moderate-length codes, very good frame error-rate performance can be achieved. For these codes, working at about 80% of the estimated capacity, good performance was achieved at lower synchronization error probabilities. For example, to ensure that at most one in a million vertices is unrecoverable (BER≤$10^{-6}$) by an LDPC code (Code 1), the OSVETA has to provide a raw deletion probability of no more than $p$=0.02. From Fig. 4. we see that this is readily achieved for most of the objects when FT is less than 6. For the same BER, random choice of vertices is unusable because the required raw deletion criterion is met only for one object and for FT lower than 4. The proposed scheme handles even higher FT if the maximum BER requirement is relaxed.

We note that no the input distribution is not optimized, i.e. the inputs were uniformly distributed. We also note that the use of distribution transformers introduce additional loss in the effective rate. However, a detailed discussion of this is beyond the scope of this analysis. We also note that although effective rates of codes for binary alphabet are low, by using codes over higher alphabets superior guarantees on maximum achievable rates may be obtained. This problem is left for future research.

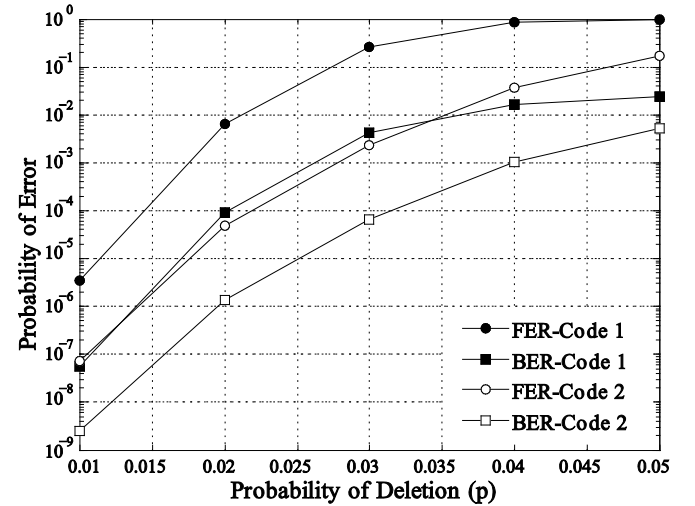

Fig. 6. Codeword and bit probabilities of error in terms of the deletion probability $p$ for the codes given in TABLE II.

*D. Malicious Attacks*

Our algorithm is designed for the non-malicious scenario in which no noise is added by a malicious user nor is 3D signal deliberately distorted in some other way in order to destroy the watermark. The method relies on the "non-continuous deletion" assumption, quantified in the "$s_d=2$" channel constraint. This is admittedly a strong assumption, but based on numerous experiments we performed, we are confident that it is valid. It is valid even when a large portion of an object is deleted. As an illustration, TABLE III shows the results for such an extreme case for the four objects considered in this section and shown in Fig. 7. For all the objects, the watermark length is 1000 bits. As it can be seen the "$s_d=2$" constraint is never violated.

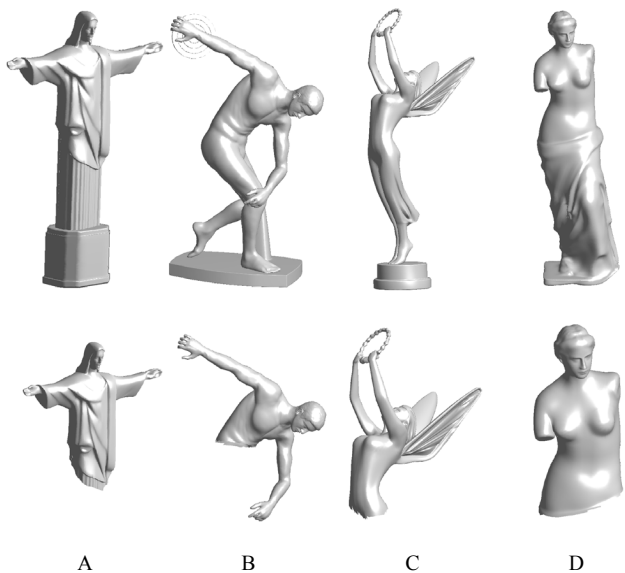

　A　　　　B　　　　C　　　　D

Fig. 7. Malicious deletion of mesh vertices. Original meshes (the first row) and maliciously damaged meshes (the second row).

TABLE III
NUMERICAL RESULTS OF MALICIOUS VERTEX DELETION.

|  | A | B | C | D |
|---|---|---|---|---|
| Total mesh vertices | 27802 | 100681 | 33465 | 17350 |
| # maliciously deleted vertices | 3893 | 77846 | 5851 | 7577 |
| watermark length | 1000 | 1000 | 1000 | 1000 |
| # consecutive deleted vertices | 0 | 0 | 0 | 0 |

However, when a malicious third party perturbs the positions of mesh vertices by adding the AWGN, with an attempt to prevent the detection of a watermark, the watermark channel introduces a combination of deletions and AWGN. To maintain the object fidelity, the malicious user must keep the noise sample amplitudes low, and as long as they remain below half a quantization step of the QIM, such jamming does not affect the watermark extraction. The fidelity and probability of bit error is thus determined by the QIM as discussed in [22]. The error correction scheme proposed in this paper is not robust to AWGN noise above the half a quantization step as the resulting bit flips affect the runlenght sequence. A bit flip within a runlength results into two shorter runs, which results in symbol insertion or/and violation of the $s_d$ constraint. Coding for a combined deletion-AWGN channel is an open problem. Very recently some progress has been reported in characterizing such channel in terms of capacity [23].

*E. Complexity*

The complexity of the proposed algorithm is determined by the complexity of its two main components: (i) the vector selection algorithm, and (ii) iterative decoding algorithm. The vector selection algorithm calculates the geometric properties of the 3D mesh (such as curvature), and involves only algebraic operations and sorting. Iterative decoding involves message updating as described in Section III.E More precisely, the complexity of iterative coding and decoding on LDPC codes is linear in code length. The decoding delay depends on the number of iterations, and a various tradeoffs are possible between quality of experience and watermark detection reliability. Typically 5-10 iterations are sufficient to achieve good performance.

## VI. DISCUSSION, CONCLUSIONS AND FUTURE WORK

We have presented a new method for 3D objects watermarking based on a combination of QIM and error correction coding. QIM is performed on the spherical coordinates of judicially selected mesh vertices stable under mesh simplification. Error correction code operates on the runlength-encoded watermark bits and ensures recovery of bits that are deleted by simplification. Runlength coding provides conceptual simplicity because it makes the watermarking process equivalent to a memoryless channel, thus allowing using powerful codes developed for these channels. The strength of the proposed watermarking coding scheme comes from carefully designed LDPC codes, while low computational complexity results from the iterative decoding algorithm used to recover deleted vertices.

In other words no noise is deliberately added by a malicious user to the 3D signal nor is signal deliberately distorted in some other way in order to destroy the watermark. Note that it is assumed that there is a cryptography "layer" on top of the data hiding layer, thus, the user data can be encrypted with desired security. These cryptographic techniques are not discussed in this paper as they operate independently from the data hiding algorithm.

Future work will include considering the attack in which a malicious third party perturbs the positions of mesh vertices by adding AWGN, with an attempt to prevent the detection of a watermark. In this case the watermark channel introduces a combination of deletions and AWGN. We are currently implementing a QIM-LDPC codes for the insertion/deletion/substitution correction. It is based on the coding scheme proposed by Davey and MacKay [3]. The probabilistic decoding algorithm operates on a two-dimensional (2D) trellis for error-correction in insertion/deletion channels. We are also implementing marker codes based on a similar decoder as proposed by Ratzer [24], and a method by Chen *at al.* [25] who combined the Davey and MacKay scheme with Varhsamov-Tenongol't block code as an inner code whose role is to detect deletions and provide the LDPC code with soft information. The above schemes, though observed to be powerful, rely on complex decoding algorithms. Therefore a study of various tradeoffs between complexity and performance of these schemes is needed.

The second direction of our future research will be the OSVETA improvements: more precise mesh geometry estimation and better curvature and topological feature estimation. We believe that these enhancements would result in more accurate identification of stable vertices and consequently significant reduction of deletion probability.

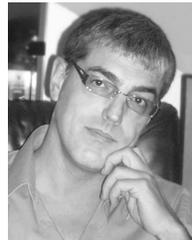

**Bata Vasic** (S'92 – M'93) received the B.Sc. and M.Sc. degrees in electrical engineering from the University of Nis, Nis, Yugoslavia (now Serbia), in 1992, and 1993, respectively. In 1992 he was a pioneer in the architectural 3D visualization in the Balkan. His research is in special effects for the movies. He produced and published more than three hundred 3D animations and TV commercials based on special effects. Since 2008 he has been teaching the Computers Animations, Computer Graphics and Computer Design courses in the Department of Electronics at Faculty of Electronic Engineering in Nis, University of Nis. His interests are in the theory of mesh models geometry in 3D space with special emphasis on using geometric features for watermark and security of 3D mesh models.

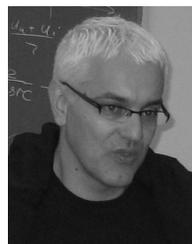

**Bane Vasic** (S'92 – M'93 – SM'02) received the B.Sc., M. Sc., and Ph.D. degrees in electrical engineering from the University of Nis, Nis, Yugoslavia (now Serbia), in 1989, 1991, and 1994, respectively. From 1996 to 1997, he worked as a Visiting Scientist at the Rochester Institute of Technology and Kodak Research, Rochester, NY, where he was involved in research in DVD copy protection. From 1998 to 2000, he was with Bell Laboratories (Bell-Labs). He was involved in research in iterative decoding and low-density parity check codes, as well as development of codes and detectors implemented in Bell-Labs chips. Presently, he is a Professor in the Electrical and Computer Engineering Department, University of Arizona, Tucson. His research interests include coding theory, information theory, communication theory, and digital communications and recording.

Dr. Vasic is a Member of the Editorial Board for the IEEE Transactions on Magnetics and the Guest Editor of the IEEE Journal of Selected Areas in Communications. He served as Technical Program Chair of the IEEE Communication Theory Workshop in 2003 and as Co-organizer of the Center for Discrete Mathematics and Theoretical Computer Science (DIMACS)Workshops on Optical/ Magnetic Recording and Optical Transmission, and Theoretical Advances in Information Recording in 2004. He was Co-organizer of the Los Alamos Workshop on Applications of Statistical Physics or Coding Theory in 2004, the Communication Theory Symposium within the IEEE International Conference on Communications (ICC 2006), and IEEE Communication Workshop (CTW 2007). He is an IEEE Fellow and daVinci Circle Fellow.